\documentstyle[11pt,doublespace]{article}

\input epsf.tex
\newcommand{\ket}[1]{{|#1\rangle}}
\newcommand{\bra}[1]{{\langle#1|}}
\newcommand{\braket}[2]{\langle#1|#2\rangle}
\newcommand{\tensor}{\otimes}
\newcommand{\isomorphic}{\simeq}
\newcommand{\circleplus}{\oplus}

\newcommand{\cA}{{\cal A}}
\newcommand{\cB}{{\cal B}}
\newcommand{\cC}{{\cal C}}
\newcommand{\cD}{{\cal D}}
\newcommand{\cE}{{\cal E}}

\newcommand{\cH}{{\cal H}}

\newcommand{\cM}{{\cal M}}

\newcommand{\cO}{{\cal O}}
\newcommand{\cP}{{\cal P}}
\newcommand{\cQ}{{\cal Q}}
\newcommand{\cR}{{\cal R}}

\newcommand{\cV}{{\cal V}}

\newtheorem{theorem}{Theorem}[section]

\newtheorem{lemma}[theorem]{Lemma}

\newtheorem{examplehidden}[theorem]{Example}

\newtheorem{definitionhidden}{Definition.}

{\end{examplehidden}\par}
{\end{definitionhidden}\par}
\newcommand{\proof}{\paragraph*{Proof.}}

\newcommand{\qed}{\hspace*{\fill}\rule{2.5mm}{2.5mm}%
\vspace*{8pt}\par}
\newsavebox{\cardofbox}
\newlength{\cardofboxheight}
\newlength{\cardofboxdepth}
\newlength{\cardoftmpld}
\sbox{\cardofbox}{$l$}
\newlength{\cardoftmpls}
\sbox{\cardofbox}{$\scriptstyle l$}
\newlength{\cardoftmplss}
\sbox{\cardofbox}{$\scriptscriptstyle l$}
\newcommand{\cardof}[1]{%
\mathchoice{%
\protect\sbox{\cardofbox}{$\displaystyle{#1}\rule{0pt}{\cardoftmpld}$}%
\setlength{\cardofboxdepth}{-\dp\cardofbox}%
\addtolength{\cardofboxdepth}{-2pt}%
\setlength{\cardofboxheight}{\dp\cardofbox}%
\addtolength{\cardofboxheight}{\ht\cardofbox}%
\addtolength{\cardofboxheight}{3pt}%
\,\rule[\cardofboxdepth]%
{.5pt}{\cardofboxheight}%
\,\usebox{\cardofbox}\,%
\rule[\cardofboxdepth]{.5pt}{\cardofboxheight}\,}%
{%
\protect\sbox{\cardofbox}{${#1}\rule{0pt}{\cardoftmpld}$}%
\setlength{\cardofboxdepth}{-\dp\cardofbox}%
\addtolength{\cardofboxdepth}{-2pt}%
\setlength{\cardofboxheight}{\dp\cardofbox}%
\addtolength{\cardofboxheight}{\ht\cardofbox}%
\addtolength{\cardofboxheight}{3pt}%
\,\rule[\cardofboxdepth]%
{.5pt}{\cardofboxheight}%
\,\usebox{\cardofbox}\,%
\rule[\cardofboxdepth]{.5pt}{\cardofboxheight}\,}
{%
\protect\sbox{\cardofbox}{$\scriptstyle{#1}\rule{0pt}{\cardoftmpls}$}%
\setlength{\cardofboxdepth}{-\dp\cardofbox}%
\addtolength{\cardofboxdepth}{-1pt}%
\setlength{\cardofboxheight}{\dp\cardofbox}%
\addtolength{\cardofboxheight}{\ht\cardofbox}%
\addtolength{\cardofboxheight}{1,5pt}%
\rule[\cardofboxdepth]%
{.5pt}{\cardofboxheight}%
\hspace*{.5pt}\usebox{\cardofbox}\hspace*{.5pt}%
\rule[\cardofboxdepth]{.5pt}{\cardofboxheight}}
{%
\protect\sbox{\cardofbox}{$\scriptscriptstyle{#1}\rule{0pt}{\cardoftmplss}$}%
\setlength{\cardofboxdepth}{-\dp\cardofbox}%
\addtolength{\cardofboxdepth}{-.5pt}%
\setlength{\cardofboxheight}{\dp\cardofbox}%
\addtolength{\cardofboxheight}{\ht\cardofbox}%
\addtolength{\cardofboxheight}{.8pt}%
\rule[\cardofboxdepth]%
{.5pt}{\cardofboxheight}%
\hspace*{.5pt}\usebox{\cardofbox}\hspace*{.5pt}%
\rule[\cardofboxdepth]{.5pt}{\cardofboxheight}}%
}

\begin{document}
\bibliographystyle{plain}

\title{A Theory of Quantum Error-Correcting Codes}

\author{
Emanuel Knill$^1$\protect\thanks{email: knill@lanl.gov}, Raymond Laflamme$^2$ \protect\thanks{laf@time.lanl.gov} 
\and
$^1$ CIC-3, MS B265, $^2$ T-6, MS B288
\and
Los Alamos National Laboratory, NM 87545, USA.
}

\date{April 1995}

\maketitle

\vskip -2.2truein\rightline{LA-UR-96-1300}
\vskip 2.truein

\begin{abstract}
Quantum Error Correction will be necessary for preserving coherent
states against noise and other unwanted interactions in quantum
computation and communication.  We develop a general theory of quantum
error correction based on encoding states into larger Hilbert spaces
subject to known interactions. 
We obtain necessary and sufficient conditions for the perfect recovery
of an encoded state after its degradation by an interaction.
The conditions depend only on the behavior of the logical states. We use
them to give a recovery operator independent definition
of error-correcting codes. We relate this definition to
four others: The existence of a left inverse of the interaction,
an explicit representation of the error syndrome using tensor products,
perfect recovery of the completely entangled state, and an
information theoretic identity.
Two notions of fidelity and error for imperfect recovery are introduced,
one for pure and the other for entangled states. The latter is more
appropriate when using codes in a quantum memory or in
applications of quantum teleportation to communication. We show that
the error for entangled states is bounded linearly by the error for
pure states. A formal definition of independent interactions for
qubits is given. This leads to lower bounds on the number of qubits
required to correct $e$ errors and a formal proof that the classical
bounds on the probability of error of $e$-error-correcting codes
applies to $e$-error-correcting quantum codes, provided that the
interaction is dominated by an identity component.
\end{abstract}

\section{Introduction}
\label{section:introduction}   

Within the past few years, quantum computation and communication have
undergone a dramatic evolution. From being subjects of primarily
academic interest, they have become fields having an enormous
potential for revolutionizing computer science and cryptography, as
well as an impact on issues of national security, and even potentially
commercializable applications.  This has resulted not only from the
development of new algorithms such as quantum factoring\cite{Shor94},
but also as a consequence of recent experimental work on
implementations of individual quantum
gates~\cite{wineland95,haroche95,turchette95} and of quantum
cryptography~\cite{hughes95}.

Unfortunately, the quantum states required to carry out a computation are
very sensitive to the imperfections of the hardware, and above all,
to the decoherence\cite{Zurek91} caused by interaction with the
environment (by environment we mean all the degrees of freedom which
can have unwanted interactions with the computer).  This fragility of a
quantum computer\cite{Landauer95,Unruh94,CLSZ95} is closely tied to its function:
it acts as a sophisticated, nonlinear interferometer.  The coherent
interference pattern between the multitude of superpositions is
essential for taking advantage of quantum parallelism, which is the key
feature allowing one to explore aspects of an exponentially large number of
possible solutions. 

To ensure that the fragility of quantum states does not destroy our
ability to extract the desired interference pattern requires
techniques for correcting errors. It is interesting to draw a parallel
between the state of the art in quantum computation today and that of
classical computers in the 40's.  At that time it was often said that
classical computers would not be very useful because errors in the
computer itself would render the result
untrustworthy~\cite{bennet95}. These doubts disappeared after the
discovery of powerful error-correction techniques. Similar doubts are
being expressed about the feasibility of the large scale application
of quantum computers. These doubts are partially based on the belief
that to perform an error-correction step, knowledge of the exact state
of the computer is required.  Such knowledge would destroy the quantum
mechanical properties of the state. However, Shor\cite{shor95} has
shown that in a restricted model of errors (similar to that which is
assumed for classical error-correction) it is possible to restore a
state using only partial knowledge of the state of the quantum
computer.  Many codes have since been discovered which correct for
specific interactions
\cite{steane95,calde95,chuang95a,LMPZ,brauns96,bennett96,vaidman96}. As a result, it may
now be possible to implement practical quantum memories and achieve
very reliable quantum communication. These ideas have opened the path
to a general theory of quantum error correction; the subject of this
paper.

This manuscript is organized as follows: In
Section~\ref{section:intuitive}, we give an intuitive approach to the
theory of quantum error correction and introduce some simple examples
of the basic concepts.  These concepts are formalized in
Section~\ref{section:codes}, where the notions of fidelity and error
of a code are introduced.  Instead of considering explicit encoding
and decoding operators, we introduce recovery superoperators. These
operators allow us to study the most general physical processes which
can be used for error-correction.  Quantum error-correcting codes
which permit complete restoration of the encoded state can then be
characterized. We give necessary and sufficient conditions for being
able to recover the state of a system after it has evolved through a
superoperator. These conditions depend only on the subspace of the
code. Several equivalent characterizations are possible and we give
four: One based on the existence of a left inverse of the interaction
superoperator, one using the explicit representation of the coding
space as a tensor product of the code with a quantum error syndrome,
one exploiting the effect of the operators on a completely entangled
state and finally one using an information theoretic identity.  In
Section~\ref{section:recovery} we discuss several methods for
implementing the recovery operator in practice and point out that if
certain additional properties hold, the recovery operator can be
substantially simplified.  Next, in Section~\ref{section:independent}
we discuss independent interactions for strings of qubits (or other
systems). These types of interactions are the natural generalization
of classical independent errors.  After a short discussion of the
physical interpretation and relevance we give a proof that it is not
possible to obtain a one-error correcting code for one qubit using a
coding space of only four qubits. This is generalized in a theorem
about correcting $e$ errors and a characterization of
$e$-error-correcting codes.  Finally we address the important issue of
the fidelity of codes with imperfect recovery operators.  We observe
that a correct measure of fidelity must take into account any
entanglements of the state. We show that the fidelity of the recovery
of an entangled state can be bounded below in terms of the pure state
fidelity. An example is provided to show that our bound is best
possible. We end this section by proving a bound on the fidelity of
codes where one of the interaction operators is proportional to the
identity. In Section~\ref{section:conclusion} we conclude the paper
with a final summary of the results and their implications.

\section{An intuitive approach}
\label{section:intuitive}
 
Coherent quantum states are used in quantum communication and quantum
computation. Both situations involve the manipulation of states by
unitary operations where some desired information is eventually
extracted from parts of the state by measurement. Quantum
communication involves multiple parties with limited communication
capabilities and focuses more on the transmission of states over
potentially noisy channels, while quantum computation involves only
one party and focuses on the unitary transformations involved in
achieving the final state. In both cases, loss of coherence occurs
while executing the necessary operations, and
when some of the
systems are either transmitted or temporarily preserved in
memory. This loss of coherence results in a reduction of the
probability of getting the correct answer after completion of the
required operations.  For short distance communication or small scale
computations, the best way to avoid errors is to minimize this loss by
isolating the state as well as possible and improving the accuracy of
the unitary transformation used.  For larger distances and long
calculations errors in the state are inevitable and it is necessary to
devise a scheme for returning the state to the desired one.
Here we focus on the problem of preserving a coherent state subject
to unwanted interactions in a quantum memory or channel.

In classical communication and computer memories, corrupted
information can be restored by introducing redundancy, for example by
copying all or part of the information to be
preserved\cite{macsloane}.  Unfortunately, it is not possible to use a simple
redundancy scheme for quantum states, primarily because the
``no-cloning'' theorem\cite{Wootters} prevents the duplication of
quantum information.  However, it has recently been
realized\cite{shor95} that it is possible to correct a state against
certain known errors by spreading the information over many qubits
through an encoding. The goal is to find an encoding which behaves in
a specific way (described below) under evolution by the interaction
superoperator. The behavior is such that it permits recovery of the
original state.  This works only for specific types of superoperators.
In practice, error-correction schemes cannot correct all errors
perfectly but only a subset of them.  The quality of a scheme can be
evaluated by its fidelity, i.e. the overlap between the corrected
state with the wanted one.
 
An essential part of the error-correction scheme is the encoding of
the quantum information.  Consider the simplest non-trivial case of
encoding a single qubit. In this case the general state to be protected is of
the form $\ket{\Psi} = \alpha\ket{0} + \beta\ket{1}$.  The idea is to
map $\ket{\Psi}$ into a higher dimensional Hilbert space (using
ancilla qubits which are assumed to be in their $|0\rangle$ states
initially):
\begin{equation}
(\alpha\ket{0} + \beta\ket{1})\ket{000\ldots} \rightarrow
\alpha\ket{0_L} + \beta\ket{1_L}
\,.
\label{coding}
\end{equation}
This defines the code. $\ket{0_L}$ and $\ket{1_L}$ are called the
{\em logical zero} and the {\em logical one} of the qubit which we
want to preserve, respectively.  The new state in Eq.(\ref{coding})
should be such that any error induced by  an incorrect functioning
of the computer maps it into one of a family of two-dimensional
subspaces which preserve the relative coherence of the quantum
information (i.e. in each subspace, the state of the computer
should be in a tensor product state with the environment).  A
measurement is then performed which projects the state into one of
these subspaces.  The original state can  be recovered by a
unitary transformation which depends on which of these subspaces has
been observed. A fact to be established in
Section~\ref{section:recovery} is that for every error-correcting
code, the original state can be recovered by a measurement followed by
a unitary operation determined by the outcome of the measurement.

In order to find good encodings,
it is essential to understand the types of error which can occur.
We assume that the initial state is $\Psi_i$, which undergoes 
interaction with an environment.  This leaves the computer 
in the reduced density matrix
\begin{equation}
\rho_f= \$ (\ket{\Psi_i})  
\,,
\label{supop}
\end{equation}
where $\$ $ is the superoperator associated with the interaction.
In the case where the environment is not initially entangled with
the system $\rho_f$ can be written in the form
\begin{equation}
\rho_f= \sum_a A_a \rho_i A_a^\dagger.
\label{evolwitha}
\end{equation}
A choice of operators $A_a$ can be determined from an
orthonormal basis $\ket{\mu_a}$ of the environment,
the environment's initial state $\ket{e}$ and the evolution
operator $U$ of the whole system as follows:
\begin{equation}
A _a= \bra{\mu_a}  U \ket{e}
\label{adef}
\end{equation}
With $A_a$ written in this way,
it can be seen that
\begin{equation}
\sum_a A^\dagger_a A_a =I  
\,.
\label{anorm}
\end{equation}
The $A_a$ are linear operators of the Hilbert space of the system and
describe the effect of the environment.  The $A_a$ are
called {\em interaction operators}.  Any family of operators $A_a$
which satisfies Eq.(\ref{anorm}) defines a superoperator.  Note that
the choice of interaction operators is not unique, they depend on the
choice of the basis $\ket{\mu_a}$ of the environment.  Two sets of
interaction operators which differ only by this choice are physically equivalent.

If there is no prior knowledge of the interaction operators which
corrupt an encoded state, it is not possible to recover $\ket{\Psi_i}$
consistently.  However, in many physical systems the $A_a$ are of a
restricted form.  For example a reasonable approximation for systems
of qubits is that the interaction with the environment is independent
for each qubit. In this case the interaction operators are tensor
products of one-qubit interaction operators.  For small error rates,
it might also be that one of the one-qubit interaction operators, say
$A_0$, is near the identity.  One can then define the number of
errors of an interaction by counting the number of operators in the
tensor product which are not $A_0$.  If there is a sufficiently small
number of errors, it may be possible to retrieve the original state
just as for classical error-correction.
 
Necessary and sufficient conditions for recovery of the state
$\ket{\Psi_i}$  are (see Section~\ref{section:codes}):
\begin{eqnarray}
\bra{0_L}  A^{\dagger}_a A_b  \ket{1_L}  &=&0 
\,,\label{condortho}\\
\langle 0_L |  A^{\dagger}_a A_b         |0_L\rangle  &=&   
\langle 1_L |  A^{\dagger}_a A_b        |1_L\rangle  
\,,\label{condnorm}
\end{eqnarray}
The first condition states that the logical zero
and one must go to orthogonal states under any error.  The second one
implies that the length and inner products of the projections of the
corrupted logical zero and one should be the same.

A sufficient but not necessary condition is that Eq.(\ref{condnorm})
is zero if  $A_a$ and $A_b$ are different. This implies that each error
maps the initial state to  orthogonal subspaces. Obviously
this permits retrieval of the original state by projecting
on these subspaces.  The more general
Eq.(\ref{condnorm}) leaves room for two different errors to be mapped
on the same two-dimensional subspace.  This possibility is allowed by
the superposition principle of quantum mechanics but 
cannot occur in classical error-correction.
 
For realistic quantum computers only a subset of possible errors can
be corrected.  An appropriate measure of the
quality of a recovered code is the fidelity~\cite{schum95}.
Fidelity is the overlap between the final state $\rho_f$ of a system
$\rho$ and the original state $\ket{\Psi_i}$.
If the combined superoperator consisting of an interaction
with the environment followed by a recovery operation
is given by $\cA = \{A_0,\ldots\}$, then the fidelity is
\begin{equation}
 F(\ket{\Psi_i},{\cal A}) = \bra{\Psi_i}\rho_f \ket{\Psi_i} =
\sum_a \bra{\Psi_i}A_a\ket{\Psi_i} \bra{\Psi_i}A_a^\dagger\ket{\Psi_i} 
\,.
\label{fidef}
\end{equation}
It gives the probability that the final state would pass a test checking
whether it agrees with the initial state.  
As we are thinking of encoding arbitrary states,
we do not know in advance the state  that will be used. We
therefore use the  minimum fidelity (that is the worst case fidelity)
\begin{equation}
F_{\min} =  \min_{\ket{\Psi}} \bra{\Psi}\rho_f \ket{\Psi}
\,.
\label{fidmin}
\end{equation}
The best quantum code maximizes $F_{\min}$. Hereafter we will drop
the subscript $\min$ to denote the fidelity of a code.

We now turn to a simple but important example to illustrate some of
the points mentioned above.  We investigate
decoherence\cite{Zurek91}, i.e. the randomization of the phase of
the initial state $|\Psi_i\rangle$. The effect of
decoherence is to decrease the size of the diagonal element of the
density matrix in a basis determined by the interaction Hamiltonian
with the environment. For one qubit, decoherence takes the
form
\begin{equation}
\ket{\Psi_i} =\alpha\ket{0} + \beta\ket{1}  \rightarrow 
\rho 
    \left ( \begin{array}{cc} \alpha\alpha^* & \alpha\beta^* e^{-\gamma} \\
                            \alpha^*\beta e^{-\gamma} & \beta\beta^*
             \end{array}
    \right )
\,,
\label{decoherence}
\end{equation}
where $e^{-\gamma}$ ($\gamma \geq 0$) parameterizes the amount of decoherence.
Decoherence can be understood in terms of the following interaction with 
the environment  
\begin{eqnarray}
\ket{e}\ket{0}&\rightarrow \ket{e_0}\ket{0} \, \nonumber\\
\ket{e}\ket{1}&\rightarrow \ket{e_1}\ket{1} 
\,.
\label{interaction}
\end{eqnarray}
with $\braket{e_0}{e_1} =e^{-\gamma}$.  Using the environment basis
$\ket{\mu_0}=\ket{e_0}$ and 
$\ket{\mu_1}= 
(\ket{e_1} - e^{-\gamma}\ket{e_0})/\sqrt{1-e^{-2\gamma}}$
we obtain the interaction operators
\begin{equation}
A_0 = \left(  \matrix{1&0\cr 0& e^{-\gamma}}
      \right)  \ \ ; \ \
A_1 = \left(  \matrix{0&0\cr 0& \sqrt{1-e^{-2\gamma}}}
      \right) 
\,.
\label{a01}
\end{equation} 

For a single qubit which is corrupted by decoherence the
minimum fidelity can be seen to be given by
\begin{equation}
F = \frac{1+e^{-\gamma}}{2} \sim 1-\frac{\gamma}{2} + \dots
\label{fiddeconebit}
\end{equation}
where the last approximation is valid for small $\gamma$.

In what follows we assume that the different qubits
have independent environments (a physically reasonable approximation)
so that the interaction operators are tensor products
of the ones given in Eq.(\ref{a01}).

A one-qubit code to correct this type of error by using three
qubits
has been devised in ref.\cite{shor95,steane95}.  To understand how it works,
it is better to change the basis state of the environment
to
$\ket{\mu_+}=(\ket{e_0}+\ket{e_1})/\sqrt{2(1+e^{-\gamma})}$ and 
$\ket{\mu_-}=(\ket{e_0}-\ket{e_1})/\sqrt{2(1-e^{-\gamma})}$.  This
gives the one qubit interaction operators
\begin{equation}
A_+ = a_+\left(  \matrix{1&0\cr 0& 1\cr}
      \right)  \ \ ; \ \
A_- = a_-\left(  \matrix{1&0\cr 0&-1\cr}
      \right) 
\,.
\label{a+-} 
\end{equation}
where $a_+=\sqrt{(1+e^{-\gamma})/2}$ and
$a_-=\sqrt{(1-e^{-\gamma})/2}$.  In this basis, the effect of the
environment is either to leave the system alone or flip the sign if
the qubit is in the state $\ket{1}$.  The encoding has the form
\begin{eqnarray}
\ket{0_L} &=& (\ket{0}+\ket{1})(\ket{0}+\ket{1})(\ket{0}+\ket{1})
\, \, \nonumber \\
\ket{1_L} &=& (\ket{0}-\ket{1})(\ket{0}-\ket{1})(\ket{0}-\ket{1})
\,.
\label{deccode}
\end{eqnarray}
This code is such that if one qubit is corrupted by the environment,
then it is possible to detect it by using a majority rule.

Assuming at most one incorrect qubit, the interaction with the
environment maps the initial state to one of the following possibilities:
\begin{eqnarray}
A_+ \ket{0_L} &=&    a_+^{3/2}\,
 (\ket{0}+\ket{1})(\ket{0}+\ket{1})(\ket{0}+\ket{1})\ \nonumber \\
A_-^1 \ket{0_L} &=& a_+^2 a_-^{1/2} \,
 (\ket{0}-\ket{1})(\ket{0}+\ket{1})(\ket{0}+\ket{1})\ \nonumber \\
A_-^2 \ket{0_L} &=&  a_+^2 a_-^{1/2} \,
 (\ket{0}+\ket{1})(\ket{0}-\ket{1})(\ket{0}+\ket{1})\ \nonumber \\
A_-^3 \ket{0_L} &=&  a_+^2 a_-^{1/2} \,
 (\ket{0}+\ket{1})(\ket{0}+\ket{1})(\ket{0}-\ket{1}) 
\,,
\label{decmap}
\end{eqnarray}
where the superscripts on the operator $A_-$ indicate which qubit is being
affected.
A similar results applies to $\ket{1_L}$.  The recovery operator is
the superoperator determined by the interactions
\begin{eqnarray}
R_+&=&     ( \ket{0_L}\bra{0_L} + \ket{1_L}\bra{1_L} )\ \nonumber \\
R_-^1&=&   ( \ket{0_L}\bra{0_L} + \ket{1_L}\bra{1_L} )\sigma_z^1\ \nonumber \\
R_-^2&=&   ( \ket{0_L}\bra{0_L} + \ket{1_L}\bra{1_L} )\sigma_z^2\ \nonumber \\
R_-^3&=&   ( \ket{0_L}\bra{0_L} + \ket{1_L}\bra{1_L} )\sigma_z^3\           
\,,
\label{recdec}
\end{eqnarray} 
where $\sigma_z^r$ is the $z$ Pauli matrix for the $r$'th qubit.
In practice the recovery operator is implemented by first performing
a measurement to determine which
error has occurred.  This can be achieved by using a series of 
controlled-not gates and measurements (with the possible involvement of ancilla qubits).
The measurements establish the relative signs in Eq.(\ref{decmap}).
Note that these relative signs are the same for the logical zero and
one after the same operator has acted and therefore the measurements
collapse the system to two-dimensional subspaces.  Once the
measurements reveal which subspace has actually occurred, it is
straightforward to recover the initial state with an appropriate
unitary transformation.

It is important to realize that this code corrects perfectly only if
at most one error occurs.  In general however, decoherence can induce
more than one error (as can be deduced from the fact that the $A_a$ in
Eq.(\ref{decmap}) do not form a superoperator).  As long as the
decoherence is small ({\em i.e.}, $\gamma$ is small), the
probability of having two or more errors will be much smaller than that of
having one error.  The minimum
fidelity can be bounded below by
\begin{equation}
F = 1- (a_-^{3} + 3a_-^2a_+) \approx  1-\frac{3}{4}{\gamma^2} +\dots
\label{eq:fidwerr}
\end{equation}
This scheme is thus an improvement over the single qubit evolution for
a small enough $\gamma$.  Using a $2n+1$ bit generalization of the code in
Eq.(\ref{deccode}), it is possible to have
fidelity be given by $1-O(\gamma^{n+1})$ for small $\gamma$, but
with a potentially large hidden constant.

\section{Quantum error-correcting codes}  
\label{section:codes}

\subsection{Fundamentals of quantum error-correcting codes}

It is now time to give a formal treatment of quantum codes.  We want
to preserve a $k$-dimensional subspace against some known errors.
This is accomplished by mapping the states into a larger,
$n$-dimensional Hilbert space.  First, let us define an
$(n,k)$-{\em quantum code} as a $k$-dimensional subspace of an
$n$-dimensional Hilbert space. The latter is called the {\em coding
space} and denoted by $\cH$. The symbol $\cC$ is used for the code.  An
{\em encoding operator\/} for ${\cC}$ is a unitary operator $E$ from a
$k$-dimensional Hilbert space ${\cQ}$ onto ${\cC}$. A {\em decoding
operator\/} is a right inverse of an encoding operator.

In practice, $k$ and $n$ are often powers of two, $k= 2^d$ and $n=2^r$
with $d<r$, and ${\cQ}$ and ${\cH}$ are tensor products of qubits. The
encoding operator can be implemented as a unitary operator on
$Q^{\tensor d}\tensor Q^{\tensor r-d}\tensor Q^{\tensor a}$, where the
last factor has $a$ ancillary qubits whose state before and after the
operation is intended to be $\ket{0}$.  The ancillas can be used as
scratch pad memory during the process of measurement needed to recover
${\cC}$.  In this case, the space ${\cQ}$ to be encoded is a
``standard'' subspace of the coding space, and the encoding operator
maps it to the intended code. Note that there are many encoding
operators which have the same effect on ${\cQ}$. This is because the
encoding defines only a part of the unitary transformation
needed. Which choice is actually used depends on efficiency
(e.g. the number of gates in a physical situation) as well
as the desired error-correcting properties.

For the purpose of discussing error-correcting properties of codes,
instead of focusing on encoding and decoding operators, we introduce
the recovery superoperator.  A {\em recovery (super)operator\/}
${\cR}$ is a superoperator on the coding space. A recovery operator is
used to restore a state to the code after it has been affected by an
interaction with the environment.  Note that except for their intended
use, recovery and interaction operators are the same type of object.

Use of a recovery operator instead of an explicit unitary operator
allows us to ignore many of the details of implementing a code which
are not relevant to its error-correcting properties.
It is general enough to represent potentially unintended or
unavoidable side-effects of the more traditional decode/encode
operations.  In practice, a recovery operator may be implemented by a
combination of unitary operations and classical measurements or
by unitary operations alone.

A {\em quantum error-correcting code\/} is a pair $({\cC}, {\cal R})$
consisting of a quantum code and a recovery operator. The correcting
properties of an error-correcting code depend on the interaction with
the environment. Let ${\cA}$ be a family of linear operators as
described in Eq.(\ref{evolwitha}). The fidelity of the code is
determined by the fidelity of the composition ${\cR}{\cA}$ restricted
to ${\cC}$.  The fidelity of the error-correcting code is thus defined
as
\begin{eqnarray*}
F({\cC}, {\cR}{\cA}) &=&
    \min_{\ket{\Psi}\in{\cC}}
    F(\ket{\Psi}, {\cR}{\cA})\\
  &=&
    \min_{\ket{\Psi}{\rm \in\ }{\cC}}
    \sum_{r,a}|\bra{\Psi} R_rA_a\ket{\Psi}|^2\,,
\end{eqnarray*}
where the $R_r$ are the interaction operators for the superoperator $\cR$.
It is
useful to consider families of linear operators which do not necessarily
satisfy the superoperator constraint Eq.(\ref{anorm}). In that
case the fidelity as defined above is not correctly normalized and instead we
consider the error of the code.
The {\em error} of the code is defined as
\begin{eqnarray*}
E({\cC}, {\cR}{\cA}) &=&
    \max_{\ket{\Psi}\in{\cC}}
 \sum_{r,a}\left|\left(R_r A_a - \bra{\Psi} R_r A_a \ket{\Psi}\right)
                  \ket{\Psi}\right|^2\,.
\end{eqnarray*}
Figure 1
%\marginpar{\tiny Can we use {\tt psfig.tex} to
%include the figure and give it a name?}
gives a geometric picture
of the notion of fidelity and error of a code.
The error of the code makes sense for arbitrary families
$\cA$. For superoperators, it is given by $1-F(\cC,{\cR}{\cA})$,
which is the worst-case probability of not observing the desired
state if we were to attempt to measure it directly.

%\begin{figure}
%\vspace{3 truein}
%\begin{center}
{\ }\vskip -0.6truein
\hskip 0.4truein\epsfxsize 5. truein\epsfbox{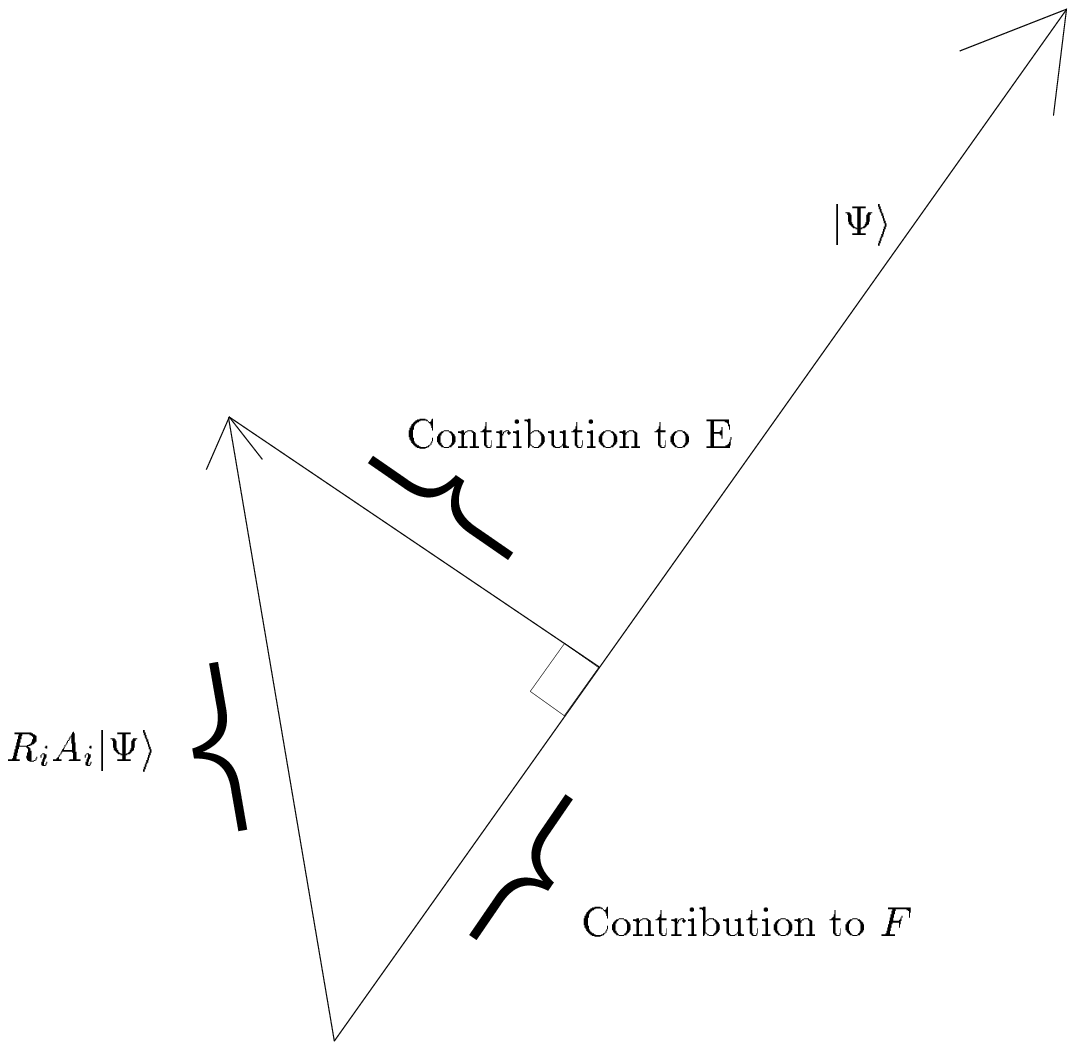}
%	\mbox{\psfig{file=figures/fig1,width=6in}}
%\end{center}
%\caption{Geometric view of the relation between fidelity and error.
%T%he fidelity is the sum of the projections along the state.
%The error is the geometric difference between the state and the fidelity.
%}
\vskip -3.2truein

\noindent Figure 1.  Geometric relation between fidelity and error.
The fidelity is the sum of the projections (for each interaction operator)
along the state.
The error gives the ``distance'' from the original state for each interaction operator.
\bigskip

We first focus on the ideal case where the code corrects all errors,
i.e. when the initial state is recovered perfectly for all operators
in $\cA$.  The case of imperfect recovery will be discussed later.
The pair $({\cC}, {\cR})$ is an {\em ${\cA}$-correcting code\/} if
$E({\cC}, {\cR}{\cA}) = 0$. Note that this is equivalent to saying
that for each $A_a$, $E({\cC}, {\cR}A_i) = 0$. Thus we can speak
of ${\cA}$-correcting codes even if ${\cA}$ is not finite. In the next
subsection we use characterizations of ${\cA}$-correcting codes
to slightly modify this definition by omitting explicit mention of the
recovery operator.

Before we characterize $\cA$-correcting codes, let us turn
the problem around and ask
what the family ${\cA}({\cC}, {\cR})$ of operators $A$ for which
$({\cC}, {\cR})$ is $A$-correcting looks like. The next result
gives an answer.  

\begin{theorem}
\label{theorem:linear}
The operator $A_a$ is in ${\cA}({\cC}, {\cR})$ iff when restricted to
${\cC}$, $R_r A_a =  \lambda_{ra} I$ for each $R_r\in{\cR}$.  The
family ${\cA}({\cC}, {\cR})$ is linearly closed
and $(\cC,\cR)$ is $\cA(\cC,\cR)$ correcting.
\end{theorem}

\proof
To be $A_a$-correcting requires that for $\ket{\Psi}\in{\cC}$,
\[
\left|\left(R_r A_a   - (\bra{\Psi} R_r A_a \ket{\Psi})\right)
      \ket{\Psi}\right| = 0.\]
This implies that $R_r A_a\ket{\Psi}=\lambda_{ra}(\ket{\Psi})
\ket{\Psi}$.  By linearity of $R_r A_a$, $\lambda_{ra}(\ket{\Psi})$
cannot depend on $\ket{\Psi}$. The rest of the theorem is immediate.
\qed

\subsection{Characterizations of ${\cA}$-correcting Codes}

So far we have defined ${\cA}$-correcting codes both in terms of
the code and the recovery operator. One of the most important
consequences of the characterizations of ${\cA}$-correcting codes
below is to allow defining ${\cA}$-correcting codes without
reference to the recovery operator. Let
$\ket{i_L}$ denote the elements of an orthonormal basis of the code
${\cC}$.  The first characterization has proved the most useful so
far for finding good codes by systematic searches
such as that in~\cite{LMPZ} or by exploiting
linear techniques from the
classical theory of error-correcting codes~\cite{steane95,calde95}.

\begin{theorem}
\label{theorem:characterization1}
The code ${\cC}$ can be extended to an ${\cA}$-correcting code
iff for all basis elements $\ket{i_L}$, $\ket{j_L}$ ($i\not= j$) and operators
$A_a$, $A_b$ in ${\cA}$
\begin{eqnarray}
\bra{i_L}A_a^\dagger A_b\ket{i_L} &=& \bra{j_L} A_a^\dagger A_b\ket{j_L}
\label{eqn:char1}
\end{eqnarray}
and
\begin{eqnarray}
\bra{i_L}A_a^\dagger A_b\ket{j_L} &=& 0.
\label{eqn:char2}
\end{eqnarray}
\end{theorem}

These conditions are more general than the ones given in~\cite{ekert96}
which are sufficient but not necessary. Since they are independent
of a recovery operator, we can define an $\cA$-correcting code
as one which satisfies Eq.(\ref{eqn:char1}) and Eq.(\ref{eqn:char2})
for any one (and therefore every) basis of the code.

\proof
Assume that $(\cC,\cR)$ is an $\cA$-correcting code.
We compute $\bra{i_L}A_a^\dagger A_b\ket{j_L}$ explicitly.
\begin{eqnarray*}
\bra{i_L}A_a^\dagger A_b\ket{j_L} &=&
    \bra{i_L}A_a^\dagger I A_b\ket{j_L}\\
  &=&
    \bra{i_L}A_a^\dagger \sum_r R_r^\dagger R_r A_b\ket{j_L}\\
  &=&
    \sum_r \bra{i_L}A_a^\dagger R_r^\dagger R_r A_b\ket{j_L}\\
  &=&
    \sum_r \bra{i_L}\bar\lambda_{ar}\lambda_{br}\ket{j_L}\\
  &=& \alpha_{ab}\delta_{ij}
\end{eqnarray*}
where we have used the superoperator properties
of $\cR$ and Theorem~\ref{theorem:linear}.
The forward direction of the theorem now follows by inspection.

Let us now show how to construct a recovery operator given that
Eq.(\ref{eqn:char1}) and Eq.(\ref{eqn:char2}) hold.  Call ${\cV}^i$
the subspace spanned by $A_a\ket{i_L}$ (for all $a$).  By
Eq.(\ref{eqn:char2}), the $\cV_i$ are orthogonal subspaces.  Let
$\ket{\nu^i_r}$ be an orthonormal basis for ${\cV}^i$.  We shall
shortly impose additional conditions on the $\ket{\nu^i_r}$. For now,
observe that the $\ket{\nu^i_r}$ are mutually orthogonal. Hence there
exist unitary $V_r$ which return $\ket{\nu^i_r}$ to the corresponding
state $|i_L\rangle$:
\begin{equation}
V_r \ket{\nu^i_r} = \ket{i_L}
\,.
\label{nutoi}
\end{equation}
The recovery operator is given by the interaction operators
\begin{equation}
{\cal R} = \{{\cal O},  R_1, \ldots,  R_r,\ldots \}
\,,
\label{recovop}
\end{equation} 
where ${\cal O}$ is the projection onto the orthogonal complement of
$\bigoplus_i {\cV}^i$, i.e. the part of the Hilbert space which is not
reached by acting on the code with the $A_a$, and
\begin{eqnarray}
R_r &=&  V_r \sum_i \ket{\nu^i_r}\bra{\nu^i_r}. 
\label{arecovop}
\end{eqnarray} 
That ${\cR}$ is a superoperator follows
from the observation that it is a sum of orthogonal projections
followed by unitary operators where the projections
span the Hilbert space.

To ensure that $\cR$ recovers the state, we need unitary operators
$U_i$ such that $U_i\ket{\nu^0_r} = \ket{\nu^i_r}$ and for all $A_a$,
$U_i A_a\ket{0_L} = A_a\ket{i_L}$.  The existence of unitary operators
satisfying the second condition follows from Eq.(\ref{eqn:char1}),
according to which the innerproduct relationships between the
$A_a\ket{0_L}$ and the $A_a\ket{i_L}$ are
identical~\cite{adaeqbdb}. Given such $U_i$, $\ket{\nu^i_r}$ can
be made to satisfy the remaining condition by choosing the basis
$\ket{\nu^0_r}$ of $\cV^0$ and {\em defining} $\ket{\nu^i_r} =
U_i\ket{\nu^0_r}$.

We show that $\cR$ does indeed recover the state, i.e. for
$\Psi\in\cC$, $R_r A_a \ket{\Psi}$ is proportional to $\Psi$.  
We can write
\begin{eqnarray}
A_a\ket{\Psi} &=& A_a\sum_i\alpha_i\ket{i_L} \nonumber \\
              &=& \sum_i\alpha_i A_a\ket{i_L}\nonumber\\
              &=& \sum_i\alpha_i U_i A_a \ket{0_L} \nonumber \\
              &=& \sum_{i,r}\alpha_i U_i \beta_{ar}^{0} \ket{\nu^0_r}  \nonumber\\
              &=& \sum_{i,r}\alpha_i\beta_{ar}^{0}\ket{\nu^i_r}
\,,     
\end{eqnarray}
where the identities define $\alpha_i$ and $\beta_{ar}^{0}$
by expansion in terms of the corresponding basis elements.
The introduction of the operators $U_i$ is what allows us to 
obtain the expansion in the last line where the $\beta$'s show
no dependence on $i$.
We can now compute $R_rA_a\ket{\Psi}$ as
\begin{eqnarray}
R_r A_a \ket{\Psi} &=& \sum_i V_r \ket{\nu^i_r}\bra{\nu^i_r}
                      \sum_{j,s}\alpha_j \beta_{as}^{0} \ket{\nu^j_s}
                      \nonumber \\
                   &=&\sum_i \alpha_i \beta_{ar}^{0}
                      V_r \ket{\nu^i_r} \nonumber \\ 
                   &=&\sum_i
                       \beta_{ar}^{0} \alpha_i\ket{i_L} \nonumber \\
                   &=&\beta_{ar}^{0} \ket{\Psi}
\,.
\end{eqnarray}
This implies that $R_rA_a$ is a multiple of the identity operation
on $\cC$. Since $\cO$ is null on all $A_a\ket{j_L}$, the fact
that $\cR$ is a recovery operator for $\cA$ follows.
\qed

An interesting observation about Eq.(\ref{eqn:char1}) is that it does
not require that the logical states have zero inner products when two
different interactions are applied, but merely that the scalar products
are the same.  For two-dimensional codes, this means that parts of the
subspaces spanned by $A_a\ket{0_L}$ and $A_a\ket{1_L}$ to which the
states are mapped may overlap. If we identify each $A_a$ with a
distinct error, then this possibility allows the correction of more
than one error per two-dimensional subspace. This is a novel feature
of quantum error-correcting codes which does not exist in their
classical counterparts.  The fact that non-trivial overlap is possible
is demonstrated by the following example.

Let us consider the code $\{\ket{0_L} = \ket{00},\ket{1_L} = \ket{11}\}$
subject to the interaction operators
\begin{equation}
A_0=\left ( \matrix{\sqrt{1-2q} &0&0&0\cr
                     0&1&0&0\cr 0&0&1&0 \cr 0&0&0&\sqrt{1-2q} \cr} \right) \,
A_1=\left ( \matrix{\sqrt{q/2} &0&0&0\cr
                     0&0&0&\sqrt{q/2}\cr 
                      \sqrt{q/2}&0&0&0 \cr 0&0&0&\sqrt{q/2}) \cr} \right) 
\nonumber
\end{equation}
\begin{equation}
A_2=\left ( \matrix{\sqrt{q/2} &0&0&0\cr
                     0&0&0&\sqrt{q/2}\cr 
                      -\sqrt{q/2}&0&0&0 \cr 0&0&0&-\sqrt{q/2}) \cr} \right) 
\,,
\label{nonortho1}
\end{equation}
for some fixed $0< q< 1$.
It is easy to check that these operators
form a superoperator.  They are linearly independent and therefore
cannot be reduced to a smaller, equivalent interaction.
The $A_i$ map the logical states as follows:
\begin{eqnarray}
\ket{0_L} &\rightarrow& \sqrt{1-2q}\ket{00} \, , \, 
                     \sqrt{q/2}(\ket{00} + \ket{10}) \, , \,
                     \sqrt{q/2}(\ket{00} - \ket{10}) \nonumber \\
\ket{1_L} &\rightarrow& \sqrt{1-2q}\ket{11} \, , \, 
                     \sqrt{q/2}(\ket{01} + \ket{11}) \, , \,
                     \sqrt{q/2}(\ket{01} - \ket{11}) 
\,.
\label{nonortho2}
\end{eqnarray}
Naively one might expect that the states on the right
hand sides are linearly independent, but
in fact, one of them is linearly dependent
on the other two in each case.
We therefore need only two recovery operators
to retrieve the initial state. They are given
by
\begin{equation}
R_0=\ket{00}\bra{00}+\ket{11}\bra{11}\, ;\,
R_1= \ket{00}\bra{10}+\ket{11}\bra{01}
\,.
\end{equation}
Whether there are any such examples of practical significance is
under investigation.

We return to the problem of
characterizing quantum error-correcting codes.
If ${\cal A}$ is a superoperator, then a simple characterization
of ${\cal A}$-correcting codes is in terms of left invertible superoperators.

\begin{theorem}
\label{theorem:characterization2}
Let ${\cA}$ be a superoperator.
${\cC}$ is an ${\cA}$-correcting code iff the restriction of
${\cA}$ to ${\cC}$ has a left superoperator inverse.
\end{theorem}

\proof
By Theorem~\ref{theorem:linear}, $\cC$ is an $\cA$-correcting code
iff there exists a superoperator $\cR$ such that on $\cC$,
$R_rA_a = \lambda_{ra} I$ for all $r$ and $a$.
This means that $\cR\cA$ is a superoperator equivalent to the identity
(by a change of basis on the environment).
\qed

Interestingly, to check that an operator
${\cal B} = {\cal R}{\cal A}$ has error $0$ on any state, it suffices to apply
$I\tensor {\cal B}$ to a completely entangled state.  In other words,
checking that the operator ${\cal B}$ has zero error for all pure
states of a system is equivalent to checking only one state which is
completely entangled with a copy of the system.

\begin{theorem}
\label{theorem:entangled}
${\cal B}$ has error $0$ on ${\cal C}$ iff $I\tensor{\cal B}
\sum_i\ket{i_L}\ket{i_L} = \lambda \sum_i\ket{i_L}\ket{i_L}$.
\end{theorem}

The equality in the theorem is to be interpreted in terms
of state ensembles: Two state ensembles are equivalent
iff they induce the same density matrix.

\proof
Let $B_r$ be a member of ${\cal B}$.
Then $I\tensor B_r$ is a member of $I\tensor{\cal B}$.
If ${\cal B}$ has error $0$ on ${\cal C}$, then
\begin{eqnarray*}
I\tensor B_r \sum_i\ket{i_L}\ket{i_L} 
  &=&
    \sum_i\ket{i_L}B_r\ket{i_L}\\
  &=&
    \sum_i\ket{i_L}\lambda_r\ket{i_L}\\
  &=&
    \lambda_r\sum_i\ket{i_L}\ket{i_L}.
\end{eqnarray*}
This implies that the ensemble $I\tensor{\cal B}\sum_i\ket{i_L}\ket{i_L}$
is equivalent to a scalar multiple of $\sum_i\ket{i_L}\ket{i_L}$.

Now suppose that the identity in the theorem holds. The fact
that the left hand side is equivalent (as a set of states) to
the right hand side implies that for each $r$,
\[
I\tensor B_r\sum_i\ket{i_L}\ket{i_L} = \lambda_r\sum_i\ket{i_L}\ket{i_L}.
\]
By applying the operator $I\tensor B_r$ to each summand and  using
the fact that the $\ket{i_L}\ket{i_L}$ are independent, this
gives $B_r\ket{i_L} = \lambda_r\ket{i_L}$. The result follows.
\nopagebreak\qed

An interesting and concise method of describing a code
which hides the recovery operator without removing it entirely
involves expressing the coding space as a sum of two
terms, the first of which is a tensor product of the code with
another space. As we will see, this perspective has several interesting
consequences. One of these consequences is the explicit distinction
between correctable versus detectable errors.

\begin{theorem}
\label{theorem:characterization3}
$\cC$ is an $\cA$-correcting code iff there is an isomorphism
$\sigma:\cH\isomorphic \cC\tensor\cE\circleplus\cD$
such that for all $A_a\in\cA$ and
$\ket{\Psi}\in{\cC}$, $A_a \ket{\Psi} = \sigma(\ket{\Psi}\tensor 
\ket{{\cal E}(a)})$
for some vector $\ket{{\cal E}(a)}$ depending on $A_a$ alone.
\end{theorem}

The idea is to ensure that the effect of the environment is clearly
separated from the state to be preserved. Thus $\cE$ takes up all the
information from the environment and the final state in $\cE$ encodes
the environment's effect on the code.
The final state in $\cE$ is called the {\em error syndrome\/}.
$\cD$ is the summand of ${\cH}$ which is
normally never reached by $\cA$, but which can be used for error
detection if so desired.  A {\em perfect\/} quantum code is one for
which $\cD$ is empty and the $\ket{{\cal E}(a)}$ span $\cE$.  Note
that in many cases of interest, a multiple of the identity map is in
$\cA$ (given by $A_0$ for example). In this case, $\cC =
\sigma(\cC\tensor \ket{{\cal E}(0)})$.

\proof
Let $\cC$ be an $\cA$-correcting code in $\cH$.  We use the notation
from the proof of Theorem~\ref{theorem:characterization1}.  Let $\cD$
be the orthogonal complement of the subspace spanned by the 
$\ket{\nu^i_{r}}$. Let $\cE$ be the Hilbert space spanned
by $\{\ket{\nu^0_r}\}_r$.  The isomorphism between $\cH$ and
$\cC\tensor\cE\circleplus\cD$ is established by letting
$\sigma(\ket{i_L}\ket{\nu^0_r}) = \ket{\nu^i_{r}}$ and defining
$\sigma$ to be the identity map on $\cD$. Let $A_a\in\cA$ and
$\ket{\Psi} = \sum_j\alpha_j\ket{j_L}\in\cC$.  Write $A_a\ket{0_L} =
\sum_r\beta^{0}_{ra} \ket{\nu^0_r}$. Applying the properties discussed
in the proof of Theorem~\ref{theorem:characterization1} gives
\begin{eqnarray*}
A_a\ket{\Psi} &=&
    \sum_{jr}\alpha_j\beta^{0}_{ar}\ket{\nu^j_r}\\
  &=&
    \sigma(\sum_j\alpha_j\ket{j_L}\tensor\sum_r\beta^{0}_{ar}\ket{\nu^0_r})\\
  &=&
    \sigma(\ket{\Psi}\tensor\sum_r\beta^{0}_{ra}\ket{\nu^0_r}).
\end{eqnarray*}
Thus we can let $\ket{{\cal E}(a)} = \sum_r\beta^0_{ar}\ket{\nu^0_r}$ to prove
the ``only if'' part of the theorem.

For the other direction we show how to construct a recovery
operator which restores the code after action of $\cA$.
Let $\ket{\nu^0_r}$ be a basis of $\cE$ and
let $R_r$ be the projection onto $\sigma(\cC\tensor\ket{\nu^0_r})$
followed by a unitary operator which maps
$\sigma(\ket{i_L}\tensor\ket{\nu^0_r})$ to
$\ket{i_L}$. Let ${\cal O}$ be the projection onto $\sigma(\cD)$.
Then the conditions on the $A_a$ imply that
$R_r A_a$ is a scalar multiple of the identity, which
gives the desired result.
\qed

Finally we mention that for superoperators $\cA$, there is a simple
information theoretic characterization of $\cA$-correcting codes due
to Nielsen and Schumacher~\cite{nielsen96}. Let $\ket{e} =
{1\over\sqrt{k}}\sum_i\ket{i_L}\ket{i_L}$ be the perfectly entangled
state of the code from which we can define the density matrices:
\begin{equation}
\bar\rho ={1\over k}\sum_{ai}  A_a\ket{i_L}\bra{i_L} A^\dagger_a
 \ \  {\rm and}\ \ 
\rho=\sum_a  I\tensor A_a\ket{e} \bra{e} A_a^\dagger\tensor I
\,.
\end{equation}
The entropy of a density matrix $\sigma$ is denoted by $S(\sigma)$.

\begin{theorem}
\label{theorem:characterization4}
Let $\cA$ be a superoperator. Then ${\cal C}$  is an $\cA$-correcting code iff
$S(\bar\rho) - S(\rho) = \log k$.
\end{theorem}

The quantity $S(\bar\rho)-S(\rho)$ is introduced as a natural notion
of mutual information in~\cite{nielsen96}.
The proof of the theorem can be found there.

\section{Implementing Recovery Operators}
\label{section:recovery}

Let us begin by observing that the recovery operator
constructed in Theorem~\ref{theorem:characterization1} consists
only of projections followed by unitary operators conditional
on the result of the projections. Implementing
such an operator is conceptually straightforward: First
you perform a measurement corresponding to the set of projections,
then, depending on the outcome of the measurement, you perform
an appropriate unitary operation. However, in quantum computation,
it is customary to assume that direct measurements can only be performed
in a standard basis of each system. This means that 
a suitable unitary transformations must be applied first in order
to rotate the measurement subspaces.

To discuss various methods for implementing the recovery
operator we need the notion of a unitary extension. Let
$W = \sum_iV_iP_i$, where the $P_i$ are orthogonal projections,
and $P_j^\dagger V_j^\dagger V_i P_i = 0$ for $i\not=j$.
Then a {\em unitary extension\/} of $W$ is any unitary $W'$
which agrees with $W$ on the range of the $P_i$. The conditions
ensure that $W'$ exists.

Let $\cR$ be described by the interaction operators $(U_0P_0, \dots,
U_{r_m} P_{r_m})$, where the $P_r$ are projections onto the orthogonal
subspaces $\cP_r$, and the $U_r$ are unitary.  Let $\cM$ be a separate
(ancillary) system with standard basis $\ket{r_M}$. Let $V_r$ be a
unitary operator on $\cM$ with the property that $V_r\ket{0_M} =
\ket{r_M}$ (i.e. $V_r$ is a unitary extension of
$\ket{r_M}\bra{0_M}$).  The operator $V = \sum_r P_r\tensor V_r$ is
unitary and has the property that $\cP_r\tensor\ket{0_M}$ goes to
$\cP_r\tensor\ket{r_M}$. (This is a generalization of the standard
controlled-not operations in quantum computing.) If $\cM$ starts
in the state $\ket{0_M}$, then we can perform $\cR$ by first applying
$V$, then measuring $\cM$ in the standard basis and finally applying
$U_r$ to the coding space if the outcome of the measurement is
$\ket{r_M}$. This is in fact the implementation of the recovery
operator suggested in~\cite{shor95,steane95}.
If it is necessary to represent the recovery
operator by unitary operators without measurement, then the
measurement and the final rotation step can be replaced by application
of the unitary operator $\sum_rU_r\tensor\ket{r_M}\bra{r_M}$.  
However, note that with this procedure, the information about the
environment's interaction with the coding space is transferred
completely to $\cM$. The only effective way in which $\cM$ can be
reused for subsequent operations is to dissipate that information
by a measurement.

Usually when using a code, there will be a time when
it is desirable to decode the state into a separate system $\cC'$
of the same dimension as $\cC$ with standard basis $\ket{i}$. The purpose
of decoding the state in this fashion may be to measure it,
or to perform unitary operations which cannot
easily be applied in the coding space directly, or
as the first step in a recovery operation where the second step
is to re-encode the state. Given an implementation of
the recovery operator, one can perform this decoding
by following the recovery operator
with the application of a unitary extension of the operator
$\sum_i\ket{0_L}\bra{i_L}\tensor\ket{i}\bra{0}$ to
$\cH\tensor\ket{0}$. This in effect swaps the state from $\cC$
to $\cC'$ after recovery.  

Here is a potentially useful method for decoding without
use of ancillas. We use the notation from
Theorem~\ref{theorem:characterization1}.  Let $Q_i$ be the projection
onto $\cV^i$.  First apply a unitary extension of $\sum_i
Q_i\tensor\ket{i}\bra{0}$ to $\ket{\psi}\tensor\ket{0}$ in
$\cH\tensor\cC'$. Then apply $\sum_iU_i^\dagger\tensor\ket{i}\bra{i}$.
Finally (if desired) measure $\cH$ to put the coding system into
a known state. As an alternative to the last unitary transformation,
one can measure $\cH$ in a special basis and follow the
measurement by a unitary operation on $\cC'$. One choice
for such a basis is given
by an arbitrary extension of the set
\[
\ket{e_{ir}} =
 \sum_j\omega^{ij}\ket{\nu_r^j},
\]
where $\omega$ is a $k$'th root of unity (we have neglected
normalization factors).  If the outcome of the measurement is
$\ket{e_{ir}}$, then the unitary transformation
$\sum_j\omega^{-ij}\ket{j}\bra{j}$ needs to be applied to $\cC'$ to
complete the decoding step.  If a $k\times k$ Hadamard matrix\cite{macsloane}  exists,
one can choose the coefficients of $\ket{\nu_r^i}$ and of
$\ket{i}\bra{i}$ to be $1$ or $-1$.

In many applications, $\cC'$ is in fact a subsystem of $\cH$,
that is $\cH = \cC'\tensor \cE'$. In that case we can
decode a state by using the isomorphism of
Theorem~\ref{theorem:characterization3}.
First identify $\cE$ with a subspace of $\cE'$
and apply a unitary extension $D$ of the operator
which takes $\sigma(\ket{i_L}\ket{a})$ to $\ket{i}\ket{a}$.
This can be followed by a measurement of $\cE$ to dissipate
the error. Note that in the case where the identity
map is corrected, such that $\cC = \sigma(\cC\tensor\ket{a_0})$,
we can apply $D^{-1}$ to $\ket{\psi}\ket{a_0}$ to
perform the encoding operation. 
Now the same circuit can be used for both encoding and decoding.
Recovery can be accomplished by applying $D$, a measurement
of $\cE$, a restoration of $\cE$ to $\ket{a_0}$ and finally
re-encoding using $D^{-1}$. The first example of such a configuration
was given in~\cite{LMPZ}.

We end this section by making a comment on codes
such as the ones suggested by Steane~\cite{steane95}
and Calderbank and Shor~\cite{calde95}.
These codes have the property that $\cH$ can
be represented as in Theorem~\ref{theorem:characterization3},
with the additional property that for a basis
$\ket{e_i}$ of $\cE$ and unitary operators $U_{ij}$,
\[
A_a\sigma(\ket{\psi}\ket{e_i})
  = \sigma(\sum_j U_{ij}\ket{\psi}\alpha_{aj}\ket{e_j})
\]
independent of $\psi$. This implies that each subspace
$\sigma(\cC\tensor\ket{e_i})$ is an $\cA$-correcting code.
This property is particularly useful in iterated applications
of the code, where recovery operators and interactions alternate.
Effectively, it suffices to project the state
after the interaction onto the subspaces $\sigma(\cC\tensor\ket{e_i}$
by using a recovery operator consisting of these projections.
The result of the projection is a correct state in an alternative
code, so it is not necessary to follow up with a unitary operator.
It is however necessary to keep track of the sequence of
outcomes of the projections, since the $U_{ij}$
change the required interpretation of the logical basis of $\cC$.

\section{Properties of codes correcting independent interactions}
\label{section:independent}

\subsection{Independent interactions}

It is difficult to discover quantum error-correcting codes for general
types of interactions.  In the classical theory of error-correction,
it is often assumed that errors occur independently for each symbol.
This assumption seems physically reasonable in many situations.  In
cases where it is not strictly true it can still lead to a systematic
approach for finding high-fidelity error-correcting codes.  We now
discuss the implications of a similar assumption for the quantum
theory. In this case, the set of symbols is replaced
by a fixed system such as the qubit. The coding space
is a tensor product of independent systems. 
To say that the interaction operator acts independently on each
component system means that it is a tensor product of single
system interactions. 
We shall focus on the case where each system is
a qubit to simplify the discussion.
Generalizations to larger systems
are straightforward.
Let $\cH=\cQ^{\tensor r} = \cQ_1\tensor\ldots\tensor\cQ_r$.
Given a one qubit superoperator $\cA$, we say
that $\cA^{\tensor r}$ acts independently on each qubit with
\[
\cA^{\tensor r} =\{A_{i_1}\tensor A_{i_2}\tensor\ldots\}_{i_1,i_2,\ldots}
\,.
\]

The assumption of independent interaction is reasonable for
the case of spontaneous emission where we can
take $\cA$ to consist of
\[
S_0 = \left(\matrix{1&0\cr 0&\sqrt{1-p^2}\cr}\right)\;,\;
S_1=\left(\matrix{0&0\cr 0&p\cr}\right).
\]
For phase randomization (decoherence) independence is a good
approximation when the effective wavelength of the environment is
smaller than the interspacing of the physical system used as qubits.
For example if the environment is modeled by a bath at finite
temperature, the condition is that the De Broglie wavelength is
smaller than the qubit's interspacing. The one-qubit
phase randomization interactions were given in Eq.(\ref{a01}).
%\[
%P_0 = \left(\matrix{\sqrt{1-p^2}&0\\0&\sqrt{1-p^2}}\right)\;,\;
%P_1 = \left(\matrix{p&0\\0&0}\right)\;,\;
%P_2 = \left(\matrix{0&0\\0&p}\right).
%\]

As in classical error-correction with fixed error-rates, it is in
general not possible to correct $\cA^{\tensor r}$ with error $0$.
And just as in the classical case, it is useful to consider
codes which correct well the ``important'' members of $\cA^{\tensor r}$,
that is those  which strongly affect only few of the qubits.
This leads to the study of $e$-error-correcting quantum codes.

An operator $A$ acting on $\cH$ is said to induce (at most) $e$ errors
if it is an $r$-fold tensor product of one-qubit operators where all
but $e$ of them are the identity.
An {\em $e$-error-correcting} code is one which
can recover from all interaction operators
inducing at most $e$ errors.

To discuss $e$-error-correction in more detail,
we need linear bases for the one-qubit
interactions.
One such
basis with the additional property that each operator is unitary
is given by
\begin{equation}
A_0 = \left(  \matrix{1&0\cr 0& 1}
      \right)   \ ; \ 
A_1 = \left(  \matrix{1&0\cr 0& -1}
      \right)   \ ; \ 
A_2 = \left(  \matrix{0&1\cr 1& 0}
      \right)   \ ; \ 
A_3 = \left(  \matrix{0&-1\cr 1& 0}
      \right) 
\,.
\label{abasis1}
\end{equation} 
These $A_a$ operators  physically correspond to: 0) leaving the system
unchanged, 1) changing the sign of the bit if it is in the $\ket{1}$ state,
2) flipping the bit 3) flipping the bit and
changing its sign if it was in the $\ket{1}$ state. 

Another useful basis
for the one qubit interactions is given by
\begin{equation}
\tilde A_0 = \left(  \matrix{1&0\cr 0& 0}
      \right)  \  ;  \
\tilde A_1 = \left(  \matrix{0&0\cr 0& 1}
      \right) \ ; \
\tilde A_2 = \left(  \matrix{0&1\cr 0& 0}
      \right) \ ; \
\tilde A_3 = \left(  \matrix{0&0\cr 1& 0}
      \right) 
\,.
\label{abasis2}
\end{equation} 
The operators $\tilde A_0$ and $\tilde A_1$ implement an ideal
measurement on the qubit. $\tilde A_2$ and $\tilde A_3$ implement
an ideal measurement followed by a bit flip.

The basis in Eq.(\ref{abasis1}) is the one used in~\cite{LMPZ} to find
the one-error-correcting five-qubit code.

\subsection{A Simple Lower Bound}

One of the simplest lower bounds on the number of classical code
words given that at least $e$ errors are to be corrected
is the Hamming bound. It is obtained by counting the number $b_e$ of
words within $e$ errors of each codeword. The product of
$b_e$ and the number of codewords cannot exceed the size of
the coding space.

For quantum codes, one can attempt a similar argument. 
Assume that we have written
the superoperator $\cA$ in a minimal form so that
each $A_a$ is independent.
In the special case where Eq.(\ref{eqn:char1}) are solved
by setting both sides to $0$, it is clear that
all states of the form $A_a\ket{i_L}$ are independent.
This implies that the total dimension of the space has to be at
least $k|\cA|$. This argument fails because
no such independence is implied by Eq.(\ref{eqn:char1})
and Eq.(\ref{eqn:char2}). One can however use
Theorem~\ref{theorem:characterization3} to see that
the total dimension has to exceed $k e$, where $e$ is
the dimension of $\cE$. If a lower bound on 
$\dim( A_0\ket{\Psi}, \dots ,A_{a_m}\ket{\Psi})$ is known, then this
is a lower bound on $e$.

As an example, consider the question of whether there are $(2^r,2)$-codes
with $r \leq 4$ qubits such that
any operator which induces at most one error can be corrected.
A natural basis for this family of operators can be derived from the
basis in Eq.(\ref{abasis1}) and consists of $1+3r$ operators.
Solving $2(1+3r)\leq 2^r$ suggests that $r$ must be at least
$5$. See~\cite{LMPZ} for an example of a code with $r=5$.  As was
pointed out in the previous paragraph, this argument
is incomplete.

We present here a different argument which proves that $r=5$ is the minimum
for one-error-correcting codes. Assume a code with $r=4$ exists.
We use the necessary and sufficient conditions given in
Eq.(\ref{eqn:char1}) and (\ref{eqn:char2}) and expand the logical zero and one as:
\begin{eqnarray}
 \ket{0_L} &=& \sum_{ijkl} \alpha_{ijkl} \ket{ijkl} \nonumber \\
 \ket{1_L} &=& \sum_{ijkl} \beta_{ijkl} \ket{ijkl} 
\,.
\label{01exp}
\end{eqnarray}
and use the interaction operators described in Eq.(\ref{abasis2})
Let us define the reduced density matrices
\begin{eqnarray}
\rho^0_{i'j'ij} &=&\sum_{kl} \alpha^*_{i'j'kl}\alpha_{ijkl} \nonumber \\
\rho^1_{i'j'ij} &=&\sum_{kl}\beta^*_{i'j'kl} \beta_{ijkl}  
\,.
\label{reddens}
\end{eqnarray}

Using those operators which induce an error on the last two qubits
in Eq.(\ref{eqn:char2}) we get
\begin{eqnarray}
 \sum_{ij}\alpha^*_{ij00}\beta_{ij00} &=&0 \nonumber\\
 \sum_{ij}\alpha^*_{ij10}\beta_{ij00} &=&0 \nonumber\\
          &\vdots& \nonumber\\
 \sum_{ij}\alpha^*_{ij11}\beta_{ij11} &=&0 
\,,  
\label{reddens1}
\end{eqnarray}
from which we conclude that the density matrices are orthogonal, i.e.
\begin{equation}
(\rho^0\rho^1)_{iji'j'} =
\sum_{klk'l'} \alpha^*_{ijkl}
       \underbrace{\sum_{i''j''}\alpha_{i''j''kl}\beta^*_{i''j''k'l'}
           }_{\mbox{$=0$ by Eq.(\ref{reddens1})}} 
       \ \ \beta_{i'j'k'l'} =0.
\,.
\label{reddensortho}
\end{equation}
On the other hand Eq.(\ref{eqn:char1}) implies that these two density
matrices are equal: Using those operators which induce an error in
the first two qubits we get
\begin{eqnarray}
\sum_{ij} \alpha^*_{00ij}\alpha_{00ij}
     &=&\sum_{ij} \beta^*_{00ij}\beta_{00ij} \nonumber\\  
\sum_{ij} \alpha^*_{10ij}\alpha_{10ij}
    &=&\sum_{ij} \beta^*_{10ij}\beta_{10ij} \nonumber\\  
   &\vdots & \nonumber\\
\sum_{ij} \alpha^*_{11ij}\alpha_{11ij}
    &=&\sum_{ij} \beta^*_{11ij}\beta_{11ij} 
\,,
\label{reddens2}
\end{eqnarray}
from which we deduce
\begin{eqnarray}
\rho^0_{iji'j'} &=& \sum_{kl} \alpha^*_{ijkl}\alpha_{i'j'kl} \nonumber\\
                &=& \sum_{kl} \beta^*_{ijkl}\beta_{i'j'kl} \nonumber \\
                &=& \rho^1_{iji'j'} 
\,.
\label{reddensequal}
\end{eqnarray}
Eq.(\ref{reddensortho}) and Eq.(\ref{reddensequal}) 
are inconsistent and imply that no such code exist.

The argument presented above can be generalized to the following
theorem: 

\begin{theorem}
A $(2^r,k)$ $e$-error-correcting quantum code must satisfy
$r\geq 4e+\lceil\log k\rceil$
\,.
\label{theorem:5min}
\end{theorem}

The task of proving this theorem is much simplified by characterizing
$e$-error correction in terms of the reduced density matrices of
the code words. Let the qubits of the coding space be labeled
by $1,\ldots,r$. For $U\subseteq \{1,\ldots, r\}$,
let $\rho(\ket{x},U)$ be the reduced density matrix of $\ket{x}$
on the qubits labeled by elements of $U$. The complement
of $U$ is denoted by $\bar U$.

\begin{theorem}
\label{theorem:echaracterization}
$\cC$ is an $e$-error-correcting code iff for all
$U\subseteq\{1,\ldots,r\}$ with $\cardof{U} = 2e$:
{\rm i)} for all $i,j$, $\rho(\ket{i_L}, U) = \rho(\ket{j_L}, U)$
and {\rm ii)} for
$i\not=j$, $\rho(\ket{i_L}, \bar U)\rho(\ket{j_L},\bar U) = 0$.
\end{theorem}

The proofs of Theorems \ref{theorem:5min} and \ref{theorem:echaracterization}
will be given elsewhere using a 
straightforward generalization of the techniques in
the earlier proof of the bound on one-error-correction.

\subsection{Relationship between the pure state and entangled state fidelity}

We have studied the recovery of corrupted states using
error-correction codes.  It is anticipated that
the states to be protected involve only a subset
of the entangled qubits of the computer or communication channel.
This means that in discussions of fidelity and error, the whole
state, not just the component being protected, must be considered.
Naturally we can compute the fidelity of a code taking
into account any part of the state not directly involved in
the interaction and recovery. The worst case fidelity for such
states is referred to as the {\em entangled state fidelity\/}
to distinguish it from the {\em pure state fidelity\/} introduced
earlier.

If the pure state fidelity
after recovery of the coded subsystem is one, then the entangled state
fidelity is one also; it does not matter if the state is
pure or if it is entangled with other systems. This observation
is invalid if we have imperfect fidelity.

\begin{theorem}
If the pure state fidelity is $F_p=1-\epsilon$, then the
entangled state fidelity is $F_e\geq 1-3\epsilon/2$.
There are examples where this bound is achieved.
\end{theorem}

\proof
We give the proof for the case where the system is two-dimensional.
We have
\begin{equation}
F_p = \min_{\ket{\Psi}\in\cC} \bra{\Psi} \rho\ket{\Psi} = 1-\epsilon
\,,
\label{fidpure}
\end{equation}
and we would like to put a bound on the entangled state fidelity 
\begin{equation}
F_e = \min_{\ket{\Psi_e}\in\cH\tensor\cC} \bra{\Psi_e} \rho_e\ket{\Psi_e}
\,.
\label{fident1}
\end{equation}
Here $\rho$ and $\rho_e$ are the final density matrix after
interaction and recovery if the initial state is $\ket{\Psi}$
and $\ket{\Psi_e}$ respectively.
Write the entangled state in the Schmidt basis as 
$\ket{\Psi_e}=\sum_i\sqrt{p_i} \ket{\psi_i^\cC}\ket{\psi_i^\cH}$
(the label $\cC$ characterizes the system on which we want to do error
correction and the label $\cH$ the system with which it is entangled).
We assume that only the system  $\cC$ is affected by an interaction
with the environment and subsequent recovery
and that the system $\cH$ has trivial dynamics.
In this case the interaction operators are tensor products of the identity
operator for the system $\cH$ and the ones given by the interactions
for the system $\cC$.  We can therefore rewrite Eq.(\ref{fident1})
as
\begin{equation}
F_e= \sum_{ij,a} p_ip_j 
   \bra{\psi_i^\cC} A_a \ket{\psi_i^\cC} 
   \bra{\psi_j^\cC} A_a^\dagger \ket{\psi_j^\cC}
\,.
\label{fident2}
\end{equation}

To obtain the bound we calculate the pure state fidelity for a superposition
of the form
$\sqrt{p_1}\psi^\cC_1 + e^{i\theta}\sqrt{p_2}\psi^\cC_2$.
Thus
\begin{eqnarray}
F_p &\leq &F(\sqrt{p_1}\psi^\cC_1 + e^{i\theta}\sqrt{p_2}
\psi^\cC_2) \hfill \nonumber \\
    &= &\sum_a \bra{\sqrt{p_1}\psi^\cC_1 + e^{i\theta}\sqrt{p_2}\psi^\cC_2} A_a 
                \ket{\sqrt{p_1}\psi^\cC_1 + e^{i\theta}\sqrt{p_2}\psi^\cC_2}
                                                   \nonumber \\
   & &\ \ \ \ \bra{\sqrt{p_1}\psi^\cC_1 + e^{i\theta}\sqrt{p_2}\psi^\cC_2} A_a^\dagger 
                \ket{\sqrt{p_1}\psi^\cC_1 + e^{i\theta}\sqrt{p_2}\psi^\cC_2}
\,.
\label{fidpure2}
\end{eqnarray}
We can now average uniformly the last equation over all values of $\theta$
to get
\begin{equation}
F_p     \leq F_e + p_1p_2 (\bra{ \psi^\cC_1}  A_a  \ket{ \psi^\cC_2}
                             \bra{ \psi^\cC_2}  A_a^\dagger \ket{ \psi^\cC_1}
                           + \bra{ \psi^\cC_2}  A_a  \ket{ \psi^\cC_1}
                             \bra{ \psi^\cC_1}  A_a^\dagger  \ket{ \psi^\cC_2})
\,.
\label{fidaver}
\end{equation}
Finally, Eq.(\ref{anorm}) puts a bound on the last term in Eq.(\ref{fidaver})
using the normalization of the interaction operator, i.e.
\begin{equation}
\sum_{i,a}\bra{\psi^\cC_i}A_a\ket{\psi^\cC_1} 
            \bra{\psi^\cC_1}A_a^\dagger\ket{\psi^\cC_i}\leq 1
\,.
\label{norma}
\end{equation}
(Note that the expression is a partial trace of a density matrix.  The
trace is partial because the interactions may take the original state
into a larger space containing $\cC$.) 
By expanding the sum over $i$
and noting that 1) the term with $i=1$ is at least $1-\epsilon$ by the
definition of pure state fidelity
and 2) all the terms are positive, we conclude that the
terms with $i\neq 1$ are bounded by $\epsilon$.  The largest achievable
value for $p_1p_2$ is 1/4.  This gives
\begin{equation}
F_e \geq 1- \frac{3\epsilon}{2}
\,.
\label{bound}
\end{equation}

For the example of decoherence in Section~\ref{section:intuitive}, it
is possible to show that $F_e = F_p$. The following example 
shows however that the bound in Eq.(\ref{bound}) can be achieved.
Consider the interaction consisting of scalar multiples of
the Pauli spin matrices,
\[
\cA =
\{{1\over\sqrt 3}\sigma_x, {1\over\sqrt 3}\sigma_y, {1\over\sqrt 3}\sigma_z\}.
\]
We show that for this example, $F(\cA) = {1\over 3}$ and
$F_e(\cA) = 0$.
Let $\ket{u} = \alpha\ket{0}+e^{i\theta}\beta\ket{1}$ with
$\alpha$ and $\beta$ real, and $\alpha^2+\beta^2 = 1$.
The fidelity of $\cA$ is obtained by maximizing the following
expression
\begin{eqnarray*}
{1\over 3}  (
 |\bra{u}\sigma_x\ket{u}|^2  &+&
   |\bra{u}\sigma_y\ket{u}|^2
 + |\bra{u}\sigma_z\ket{u}|^2  ) \\
 &=&
 {1\over 3} \left(
   (2\alpha\beta\cos(\theta))^2 
  + (2\alpha\beta\sin(\theta))^2
  + (\alpha^2 - \beta^2)^2\right)\\
 &=&
 {1\over 3} \left(
  (\alpha^2 + \beta^2)^2
  \right)\\\\
 &=& {1\over 3}.
\end{eqnarray*}
Hence $F(\cA) = {1\over 3}$.
To show that $F_e(\cA) = 0$, apply $\cA$ to the second system
of the completely entangled state
$\ket{e} = {1\over\sqrt 2}(\ket{0}\ket{0}+\ket{1}\ket{1})$.
We get
\begin{eqnarray*}
I\tensor\sigma_x\ket{e} &=&
    {1\over\sqrt 2} (
      \ket{0}\ket{1} + \ket{1}\ket{0}
    ), \\
I\tensor\sigma_y\ket{e} &=&  
    {i\over\sqrt 2} (
      \ket{0}\ket{1} - \ket{1}\ket{0}
    ),\\
I\tensor\sigma_z\ket{e} &=&
    {i\over\sqrt 2} (
      \ket{0}\ket{0} - \ket{1}\ket{1}
    ).
\end{eqnarray*}
These states are all orthogonal
to $\ket{e}$, whence $F_e(\cA) = 0$.  Thus this example achieves
equality in Eq.(\ref{bound}) and our bound is the best possible.
\qed

\subsection{Bounds on the fidelity of error-correcting codes for
independent interactions}
\label{subsection:independentbounds}

Let $\cA$ be a one qubit interaction of the form
$\cA = \{A_0, A_1,\ldots\}$ with $A_0$ close to the identity
in some sense. In this case we would hope that an $e$-error-correcting
code on $n$ qubits reduces the error after independent interactions
of each qubit with $\cA$. That this does indeed hold is an important
observation for the application of these error-correcting codes.
We are about to show
that in the case where $A_0 = \sqrt{1-p} I$, the classical bounds
on the probability of error in the corrected code do apply,
as has been informally discussed by Calderbank and Shor~\cite{calde95},
Steane~\cite{steane95} and others. When $A_0$ is not a scalar multiple
of the identity, then additional terms must be added to
the bounds. We defer the discussion of this case to future papers.

Assume then that $\cA = \{\sqrt{1-p}I, A_1,\ldots\}$. Denote
$\cA' = \{A_1,\ldots\}$ and note that the strength of $\cA'$ is
\[
|\cA|^2 = \sup_{\ket{x}}\sum_{i\geq 1}\bra{x}A_i^\dagger A_i \ket{x}
        = p.
\]
Let $\cC\subseteq \cQ^{\tensor r}$ be an $r$-qubit $e$-error-correcting
code with recovery operator $\cR$. To estimate the error
after recovering from $\cA^{\tensor r}$, write
\begin{eqnarray*}
\cA^{\tensor r} &=&
  \{\sqrt{1-p}I, \cA'\}^{\tensor r}\\
  &=&
  \sum_{0\leq k\leq r}\sum_{U\subseteq \{1,\ldots,r\}, |U|=k}
    \sqrt{1-p}^k (\tensor_{i\not\in U}I)\tensor(\tensor_{i\in U}\cA'),
\end{eqnarray*}
with the obvious interpretation of the tensor products and which
system each factor is acting on.
Let $\cA_U = (\tensor_{i\not\in U}I)\tensor(\tensor_{i\in U}\cA')$
refer to the ensemble of operators obtained by letting $I$ act on the
qubits in $U$ and $\cA'$ on the qubits not in $U$.  By the properties
of the recovery operator, for $|U|\leq e$, the error due to $\cR\cA_U$
is $0$. Thus it suffices to bound the error of the remaining terms
in the sum for the interaction. We do this by assuming
that the error in each summand is maximal. That is, the
contribution to the total error by $\cA_U$ is bounded by the strength
of $\cA_U$ given by the maximum
value of $|\cA_U\ket{x}|^2$. The strength of the tensor product
of operator ensembles can be computed using the next lemma.

\begin{lemma}
Let $\cB_1$ and $\cB_2$ be operator ensembles. Then
$|\cB_1\tensor\cB_2|^2 = |\cB_1|^2|\cB_2|^2$.
\end{lemma}

The lemma can be proved by diagonalizing 
$\cB^\dagger_1\cB_1 =\sum_i B^\dagger_{1i} B_{1i}$ and
$\cB^\dagger_2\cB_2 =\sum_i B^\dagger_{2i} B_{2i}$.

We deduce that the strength of $\cA_U$ is $p^{|U|}$.
By evaluating the sums over the $U$'s we obtain the
following result:

\begin{theorem}
Let $\cR$ be the recovery operator of an $e$-error correcting
code $\cC$ on $n$ qubits and $\cA = \{\sqrt{1-p}I,\cA'\}$ a superoperator
on one qubit. Then
\[
F(\cC,\cR\cA^{\tensor r})
\geq 1-\sum_{k>e} {r\choose k} p^k (1-p)^{r-k}.
\]
\end{theorem}

Note that for applications involving entanglements, the bound
needs to by modified in consideration of the relationship
between pure state and entangled state fidelity.

\section{Conclusion and future work}
\label{section:conclusion}

We have laid the foundations for a theory of quantum error-correcting
codes by providing a general definition of quantum codes and by
characterizing those which can correct known interactions with zero
error. The main features of our approach include treating a code
solely in terms of its subspace in a larger Hilbert space and defining
decoding operations in terms of general recovery superoperators. This
allows studying codes and their properties for arbitrary interaction
superoperator and avoids explicitly dealing with decoding and encoding
issues when studying the fidelity of a code given its recovery
operator.  The treatment in terms of interaction operators directly
leads to the characterizations of error-correcting codes given in
Section~\ref{section:codes}. The characterization in terms of how the
operators map individual states (Theorem \ref{theorem:characterization1}) has
proved particularly useful for finding new codes.

Our approach is not confined to the study of codes which
allow perfect reconstruction of the encoded states.  As an example of
what can be done, we formally defined $e$-error-correcting codes on
strings of qubits and considered the effect of independent
interactions. We showed that for interactions with an identity
component, there is a natural way in which the classical bound on the
error can be applied, as has been discussed informally by other
authors. This justifies the effort that has been put into
finding good $e$-error-correcting codes. We observe that
this classical bound may be more pessimistic than necessary,
but leave a careful study of the fidelity of various
known codes to future work. 

We brought up the important issue of how reliable a predictor
the pure state fidelity is for error propagation in entangled systems
and showed that the entangled state fidelity is not much less than
the pure state fidelity. The fact that it can be less is an important
observation, lest one be deceived into believing that a fidelity
of $1/3$ might be adequate if not compounded by other errors on
the same system.

The study of imperfect fidelity codes is far from complete. Both
the sources of introduced error, and its propagation when recovery
is attempted many times require further study. Ultimately, these
issues determine the circumstances when an advantage may be gained
from using error-correction schemes.

We would like to finish by commenting on a general issue.  
The present work on quantum error-correction assumes
that no errors are produced during operations.  This is a reasonable
assumption if the coding, recovery and decoding operations
take a small time compared to the rate at which errors appear
(i.e. the interaction strengths), and the error in the operations
themselves is small compared to the error corrected by the code.
We do not believe that this assumption will remain valid in the
context of large scale quantum calculations.
It is therefore important
to take into account the fact that operations are imperfect.
An important step in this direction
has already been taken in
\cite{chuanglaf96}.  There the particular case of correcting for
decoherence (phase randomization) using the three-bit scheme
presented in the introduction has been investigated.

\section{Acknowledgment}
We would like to thank I. Chuang, C. Miquel, J. Paz, B. Schumacher, J. Smolin and W. Zurek
for useful conversations. R.L. is grateful to J. Gregson
for insights on the intuitive approach to error correction.  
We have both benefited from interaction with the Quantum Computer group
at Los Alamos National Laboratory.  
This work was partially performed under the
auspices of the U.S. Department of Energy under Contract No. W-7405-ENG-36.

\end{document}